# Downregulation of aquaporin 3 promotes hyperosmolarity-induced apoptosis of nucleus pulposus cells through PI3K/Akt/mTOR pathway suppression

Yuan Sang [1,†], Huiqing Zhao [1,†], Jiajun Wu [1,†], Ting Zhang [2], Wenbin Xu [1], Hui Yao [1], Kaihua Liu [1], Chang Liu [1], Junbin Zhang [1], Ping Li [1], Depeng Wu [1], Yichun Xu [1], Jianying Zhang [2,*], Gang Hou [1,*]

[1]Department of Orthopaedics, The Third Affiliated Hospital of Sun Yat-sen University, Guangzhou, PR China.

[2]Department of Orthopaedic Surgery, University of Pittsburgh, Pittsburgh, USA

†These authors equally contributed

*Address for correspondence:

Dr. Gang Hou

The Third Affiliated Hospital of Sun Yat-sen University

NO. 2693 Kaichuang Road, Guangzhou, 510630, People's Republic of China

Tel.: 86-020-82179720

E-mail: hougang@mail.sysu.edu.cn

Or

Jianying Zhang, PhD

University of Pittsburgh

E1640 Biomedical Science Tower, 200 Lothrop Street, Pittsburgh, PA 15213, USA

Tel.: 412-383-5302

E-mail: jianying@pitt.edu

**Abstract**

Hyperosmolarity is a key contributor to nucleus pulposus cell (NPC) apoptosis during intervertebral disc degeneration (IVDD). Aquaporin 3 (AQP3), a membrane channel protein, regulates cellular osmotic balance by transporting water and osmolytes. Although AQP3 downregulation is associated with disc degeneration, its role in apoptosis under hyperosmotic conditions remains unclear. Here, we demonstrate that hyperosmolarity induces AQP3 depletion, suppresses the PI3K/AKT/mTOR signaling pathway, and promotes mitochondrial dysfunction and ROS accumulation in NPCs. Lentiviral overexpression of AQP3 restores this pathway, attenuates oxidative damage, and reduces apoptosis, preserving disc structure in IVDD rat models. In contrast, pharmacological inhibition of AQP3 exacerbates ECM catabolism and NP tissue loss. Our findings reveal that AQP3 deficiency under hyperosmolarity contributes to NPC apoptosis via suppression of PI3K/AKT/mTOR signaling, potentially creating a pathological cycle of disc degeneration. These results highlight AQP3 as a promising therapeutic target for IVDD.

**Keywords:** IVDD, NPCs, Hyperosmolarity, AQP3, PI3K/Akt/mTOR pathway, ROS

**Introduction**

Lower back pain (LBP), ranked as the leading global cause of disability, originates predominantly from intervertebral disc degeneration (IVDD) - a progressive degenerative condition affecting 40% of adults aged over 40. Beyond its debilitating impact on quality-adjusted life years, IVDD imposes substantial socioeconomic burdens [1]. The multifactorial pathogenesis of IVDD involves a complex interplay of cellular senescence, biomechanical stressors, inflammatory cascades, and genetic predisposition [2-5]. Despite these mechanistic insights, the precise spatiotemporal regulation of disc degeneration remains poorly understood at the molecular and cellular levels. Current therapeutic paradigms remain palliative, offering only transient relief via physical therapy, pharmacological agents, and invasive surgical procedures [6-9]. This critical gap in the availability of disease-modifying therapies highlights an unmet clinical need. Targeted molecular interventions that address the underlying pathobiology of IVDD progression are urgently required.

The intervertebral disc (IVD) is a sophisticated trilaminar structure comprising the proteoglycan-rich nucleus pulposus (NP), the collagenous annulus fibrosus (AF), and the cartilaginous endplates [10]. This architectural organization enables the IVD to act as a viscoelastic shock absorber that distributes compressive forces and preserves spinal flexibility and range of motion [11]. The hallmarkfeatures of IVDD encompass progressive cell lossand extracellular matrix (ECM) degradation, particularly within the NP compartment [12]. NPCs reside within one of the most physiologically challenging microenvironments in mammalian systems, characterized by hyperosmolarity, chronic hypoxia, low pH and excessive mechanical stress [13]. In the IVD microenvironment, the elevated osmotic pressure primarily arises mainly from a proteoglycan-rich ECM. Notably, the physiological osmotic pressure in NP tissue maintains oscillates between 450-550 milliosmoles per kilogram (mOsm/kg), which is much higher than that of conventional extracellular fluids in mammalian systems. Mounting evidence indicates that this pathological hyperosmolarity exerts profound regulatory effects on NPCs homeostasis, including metabolic activity, apoptosis resistance, and phenotypic maintenance. However, the mechanisms by which hyperosmolarity regulates NP cell responses remain unclear in IVDD [14-16]. Elucidating how hyperosmolarity regulates NP cell responses may provide critical insights into IVDD pathophysiology and unveil novel therapeutic targets.

Local osmotic pressure in the IVD fluctuates markedly. In response, resident cells deploy specific proteins and signaling pathways to adapt osmotically. The aquaporin (AQP) family has emerged as a key regulator of cellular osmoregulation in the IVD microenvironment [17]. AQPs are transmembrane channels that facilitate water transport across lipid bilayers and thus maintain cellular osmotic homeostasis [18-20]. In IVDD models, aquaporin expression in embryonic notochordal cells is responsive to osmotic gradients. Transcriptional changes then influence lineage commitment and apoptosis in murine IVD [21]. Aquaporin 3 (AQP3), an aquaglyceroporin, shows differential expression in the human NP and AF [21,22]. Proteomic analysis of degenerative discs shows significant AQP3 downregulation and impaired glycerol/water transport [23]. These data support a mechano-osmotic coupling model. AQP3-mediated flux may convert changes in matrix hydration and osmolarity into cytoskeletal remodeling during IVDD [21]. Here, we investigate the role of AQP3 in NPC fate under hyperosmotic stress. We focus on the molecular mechanisms driving osmoadaptive reprogramming.

**Materials and Methods**

*Extraction of rat NPCs*

Thirty Sprague-Dawley rats (female, 200-220 g, 12-week-old) were purchased from Guangdong Sijia Jingda Biotechnology Co., Ltd. (SCXK (Xiang) 2021-0002) and acclimatized for 7 days in specific-pathogen-free (SPF) conditions with ad libitum access to autoclaved feed and water. All procedures were performed at Guangzhou Forevergen Biosciences Medical Laboratory Animal Center following protocol approval by the Institutional Animal Care and Use Committee (SYXK (Yue) 2023-0186).

Euthanasia was performed using gradual $CO_2$ displacement. Lumbar intervertebral discs (Co6/7-Co8/9) were aseptically excised within 5 min post-mortem under laminar flow. NP tissue was microdissected from the innermost annular region under a dissecting microscope. Sequential enzymatic digestion was performed in 0.25% trypsin-EDTA (T4049, Sigma-Aldrich, USA) at 37°C for 5 min with 150 rpm orbital shaking and 0.25% collagenase type II (C2-BIOC, Sigma-Aldrich, USA) for 10 min under identical conditions. Digestion was terminated with 10% fetal bovine serum (FBS) (A3160802, Gibco, USA)-containing DMEM/F12 (11320033, Gibco, USA). Cell suspensions were filtered through 70 μm nylon mesh (352350, Corning, USA), centrifuged at 300×g for 5 min, and resuspended in complete medium: DMEM/F12 supplemented with 10% FBS, 1% penicillin-streptomycin (15140122, Gibco, USA), and 25 μg/mL ascorbic acid (A4544, Sigma-Aldrich, USA). Primary NPCs were maintained at 37°C in 5% $CO_2$ with medium changes every 48 h until 80% confluency.

*Culture of rat NPCs under different osmotic pressure environments*

NPCs were resuspended in DMEM/F12 medium containing 10% FBS and 1% penicillin-streptomycin, then maintained in a humidified incubator with standard atmospheric conditions (37°C, 21% $O_2$, and 5% $CO_2$). To establish different osmotic pressure conditions, sodium chloride (NaCl) (S805280, Macklin, China; 9.7 g/L, 16.2 g/L) was dissolved in the culture medium to achieve osmolality levels of 330 mOsm/kg (physiological level) and 550 mOsm/kg (hyperosmolarity level). The final osmolality was quantitatively confirmed through triplicate measurements using a freezing-point osmometer (FM-8P, Shanghai Medical College Instrument Co., Ltd., China).

*Detection of apoptosis in NPCs by flow cytometry*

NPCs were seeded in 6-well plate and exposed to 330 or 550 mOsm/kg media for 48 h (37°C and 5% $CO_2$). Following treatment, cells were enzymatically dissociated using 0.25% trypsin, washed twice with PBS, and collected by centrifugation at 300×g for 5 min at 4°C. Apoptosis analysis was performed using the Annexin V-FITC/PE Apoptosis Detection Kit (KGA1102, KeyGEN BioTECH, China) according to the manufacturer's protocol. Briefly, $1×10^5$ cells/mL were resuspended in 100 μL binding buffer and dual-stained with 5 μL Annexin V-FITC and 10 μL PE for 15 min at 25°C under light-protected conditions. Quantitative flow cytometric analysis was conducted using a FACSCalibur system (BD Biosciences, CA). Fluorescence signals were detected through 530/30 nm (FITC) and 585/42 nm (PE) bandpass filters, with data acquisition performed using CellQuest Pro software.

*Detection of caspase-3 activity by flow cytometry*

NPCs were plated in 6-well plate at a density of $1×10^5$ cells per well and maintained in a humidified incubator (37°C, 5% $CO_2$). When the cells reach 80-90% confluence after 48 h of culture, cell monolayers were washed twice with buffer. Caspase-3 proteolytic activity was measured using the Caspase-3 Activity Assay Kit (C1168S, Beyotime Biotechnology, China), strictly adhering to the manufacturer's protocol.

*TUNEL staining*

Apoptosis in rat NPCs was assessed using a commercial TUNEL kit (ZY81026, ZeYe, Shanghai, China). After dewaxing and rehydration, paraffin sections were processed sequentially: 100 μL proteinase K working solution was added to the NPCs section and incubated at 37°C for 20 min. The section was rinsed in 1 × PBS for 5 min × 3 times. Each section was dropped for 100 μL 1 × Equilibration Buffer to cover at room temperature for 10 min and incubated with 50 μL TdT incubation buffer at 37°C for 1 h. DAPI working solution was dyed at 37°C for 10 min and was coverslipped using buffered glycerin. The sample was imaged under a fluorescence microscope (Leica Microsystems, Germany). The number of TUNEL-positive cells (green) and total nuclei (blue) were counted in three randomly selected fields. The ratio of the number of TUNEL-positive cells to the number of total nuclei was calculated by cell counting method.

*Western blot assay*

Protein expression levels were quantified using Western blot analysis with β-actin serving as the endogenous loading control. Cellular proteins were extracted from NPCs using ice-cold radioimmunoprecipitation assay (RIPA) lysis buffer (P0013B, Beyotime Biotechnology, China). Protein concentrations were subsequently measured using a bicinchoninic acid (BCA) assay kit (23225, Thermo Fisher Scientific, USA) according to the manufacturer's protocol. Equal amounts of protein lysates were resolved through 10% sodium dodecyl sulfate-polyacrylamide gel electrophoresis (SDS-PAGE; P0012A, Beyotime Biotechnology, China) and subsequently transferred onto polyvinylidene difluoride (PVDF) membranes (IPVH00010, Merck Millipore, USA) using a semi-dry transfer system. Primary antibodies include anti-Bax (1:1000; ab32503, Abcam, UK); anti-Bcl-2 (1:1000; ab59348, Abcam, UK); anti-type II collagen (1:5000; 600-403-104, Thermo Fisher Scientific, USA); anti-MMP3 (1:1000; ab52915, Abcam, UK); anti-AQP3 (1:1000; ab307969, Abcam, UK); anti-mTOR (1:1000; 2983, Cell Signaling Technology, USA); anti-p-mTOR (1:1000; 5536, Cell Signaling Technology, USA); anti-PI3K (1:1000; 4249, Cell Signaling Technology, USA); anti-p-PI3K (1:1000; 13857, Cell Signaling Technology, USA); anti-Akt (1:2000; 2920, Cell Signaling Technology, USA); anti-p-Akt (1:2000; 4060, Cell Signaling Technology, USA); anti-β-actin (1:5000; 93473, Cell Signaling Technology, USA). Following primary antibody incubation, membranes were washed three times with Tris-buffered saline containing 0.1% Tween-20 (TBST; ST671, Beyotime Biotechnology, China) under gentle agitation. Subsequently, membranes were probed with species-matched horseradish peroxidase (HRP)-conjugated secondary antibodies (Goat anti-mouse IgG (1:2000; 7076, Cell Signaling Technology, USA) and Goat anti-rabbit IgG (1:2000; 7074, Cell Signaling Technology, USA)) at room temperature for 2 h with continuous shaking. After three additional TBST washes, immunoreactive bands were visualized using an enhanced chemiluminescence (ECL) kit (A38556, Thermo Fisher Scientific, USA). Protein band intensities were quantified through densitometric analysis with ImageJ software

(National Institutes of Health, USA), normalized to β-actin expression levels.

*Detection of ROS levels in NPCs by flow cytometry*

Intracellular reactive oxygen species (ROS) levels were quantified using the 2',7'-dichlorodihydrofluorescein diacetate (DCFDA) Cellular ROS Detection Assay Kit (ab113851, Abcam, UK), with protocol optimization. Briefly, NPCs were incubated with 10 μM DCFDA in serum-free RPMI-1640 medium at 37°C under 5% $CO_2$ workstation for precise environmental control. Following 30 min dark incubation, cells were washed twice with thermostated PBS and resuspended in phenol red-free DMEM/F12 to minimize autofluorescence. Data were acquired using BD FACSDiva software with threshold set at FSC-H 200. The 2',7'-dichlorofluorescein (DCF) mean fluorescence intensity (MFI) was quantified after spectral overlap compensation using single-stained controls. ROS levels were expressed as fold change relative to vehicle-treated controls.

*Detection of ROS levels in NPCs by immunofluorescence*

Mitochondrial ROS were quantified using a dual-fluorescence colocalization assay. NPCs cultured under osmotic pressure of 330 and 550 mOsm/kg were co-stained with MitoSOX Red mitochondrial superoxide indicator (M36008, Thermo Fisher Scientific, MA) and MitoTracker Green FM (C1048, Beyotime Biotechnology, China) according to the manufacturers' protocols with optimization. Briefly, cells were incubated with the dye mixture in Hank's Balanced Salt Solution at 37°C under 5% $CO_2$ for 30 min, followed by three washes with pre-warmed HBSS. Live-cell imaging was performed using a Leica DMi8 inverted fluorescence microscope. Image acquisition was conducted using LAS X software with identical exposure settings. Colocalization analysis was performed using ImageJ.

*Determination of GPx and MDA levels*

Malondialdehyde (MDA) concentrations and glutathione peroxidase (GPx) activity were quantified using standardized colorimetric and enzymatic assays, respectively. The MDA levels were determined via the thiobarbituric acid reactive substances (TBARS) method using a commercial kit (RK09070, ABclonal, China), while GPx activity was measured through NADPH oxidation kinetics with a kit (EEA010, Thermo

Fisher Scientific, MA). Briefly, cell lysates were centrifuged at 12,000×g for 15 min at 4°C to obtain supernatant fractions. Absorbance readings were acquired using a Synergy H1 microplate reader. The lower detection limits were validated as 0.1 μM for MDA and 5 U/mL for GPx using serial dilutions of reference standards.

*MMP assessment*

Mitochondrial depolarization was evaluated using the JC-1 assay kit (C2006, Beyotime Biotechnology, China) following established protocols. Treated NPCs were incubated with JC-1 working solution at 37°C under dark conditions for 30 min. Following triple washing with PBS, cells were resuspended in fresh complete medium and immediately analyzed using an inverted fluorescence microscope (DMi8, Leica Microsystems, Germany). Mitochondrial integrity was determined by quantifying the red/green fluorescence intensity ratio.

Quantitative MMP measurement was performed using JC-1 staining coupled with flow cytometry. Apoptosis-induced NPCs were harvested and washed with ice-cold PBS, then resuspended in 1× assay buffer containing JC-1. After 20 min incubation at 37°C in the dark, cells were centrifuged and washed twice with warm PBS. Cellular fluorescence was immediately analyzed using a FACSCalibur flow cytometer (BD Biosciences, USA).

*RNA sequencing*

RNA extraction from NPCs was performed using TRIzol (15596026, Thermo Fisher Scientific, USA), followed by determination of RNA concentration and purity by Nanodrop2000 microspectrophotometer. The integrity of RNA samples was examined by the Labchip GX touch microfluidic capillary system. The raw RNA sequencing data were subjected to quality control (QC) to assess the suitability of the sequencing data for subsequent analysis. RNA sequencing was performed on the Illumina HiSeq platform using next-generation sequencing (NGS) technology. To evaluate the functionality of candidate targets in IVDD, the Gene Ontology (GO) and Kyoto Encyclopedia of Genes and Genomes (KEGG) pathway enrichment analysis of co-expressed genes was performed using the clusterProfiler software package in R software. GO and KEGG pathways with $P < 0.05$ were considered to be significantly enriched.

*Transfection of expression vectors into NPCs AAV vector construction and transfection*

The pAAV-AQP3 expression plasmid was engineered by subcloning full-length human AQP3 (GenBank accession NM_004925) cDNA into the multiple cloning site of the pAAV-MCS vector (Agilent Technologies, USA) under control of the CMV immediate-early promoter. The insert orientation and sequence fidelity were verified through bidirectional Sanger sequencing (GENEWIZ, China) using vector-specific primers (forward: 5'-CGCAAATGGGCGGTAGGCGTG-3'; reverse: 5'-CTCAGTTGGCGAGCTCGGATC-3'). NPCs were plated at a density of $2.0×10^5$ cells/well on poly-L-lysine-coated glass coverslips (0111520, Thermo Fisher Scientific, USA) in 6-well plates (140675, Nunc, Denmark) and cultured overnight in DMEM/F-12 medium supplemented with 10% FBS. Transfection complexes were prepared by mixing 2.5 μg pAAV-AQP3 plasmid with 7.5 μL Lipofectamine 2000 (11668019, Thermo Fisher Scientific, USA) in Opti-MEM reduced serum medium (31985062, Gibco, USA), following a 3:1 lipid: DNA ratio. After 20 min incubation at room temperature, the complexes were added to cells maintained in antibiotic-free medium. Following 6 h incubation at 37°C with 5% $CO_2$, the transfection medium was replaced with complete growth medium for 48 h prior to subsequent analyses.

*AAV-mediated expression of AQP3*

Recombinant adeno-associated virus (rAAV) vectors were generated through triple-plasmid co-transfection in HEK293T cells (CRL-3216, ATCC, USA). The payload plasmid pAAV-TBG-AQP3-GFP contained human AQP3 cDNA fused with enhanced GFP reporter, hepatocyte-specific thyroxine-binding globulin (TBG) promoter and AAV2 inverted terminal repeats (ITRs). The packaging system consisted of pAAV-RC2/8 providing AAV2 replication (Rep) and AAV8 capsid (Cap) proteins and pHelper supplying adenoviral E2A, E4, and VA RNA genes.

At 72 h post-transfection, the cells were lysed with 0.5% sodium deoxycholate (D6750, Sigma-Aldrich, USA) and subjected to cesium chloride gradient ultracentrifugation (Optima XE-90, Beckman Coulter, USA) at 210,000×g for 48 h at 4°C. Viral fractions with refractive indices between 1.365-1.371 were collected and dialyzed against PBS containing 5% sorbitol (S3889, Sigma-Aldrich, USA). Genome titers ($1.2×10^{13}$ vg/mL) were determined by absolute quantification using ITR-targeted qPCR (Forward: 5'-GGAACCCCTAGTGATGGAGTT-3'; Reverse: 5'-CGGCCTCAGTGAGCGA-3') with linearized plasmid standards.

*Animal experiment*

Forty 12-week-old female Sprague-Dawley rats were randomly allocated into four experimental groups using stratified randomization based on body weight: Control, Annulus fibrosus puncture (AFP), AFP+AQP3 inhibitor DFP00173 (HY-126073, MedChemExpress, China), AFP+AAV-AQP3.

Animals were anesthetized via intraperitoneal injection of sodium pentobarbital (40 mg/kg; P3761, Sigma-Aldrich, USA) and positioned prone on a warming pad. Under fluoroscopic guidance (C-arm Ziehm Vision RFD 3D, Germany), the caudal intervertebral disc (Co6/7 level) was exposed through a posterolateral approach. A 26-gauge spinal needle (405210, B. Braun, Germany) was inserted perpendicular to the annulus fibrosus at 5 mm depth, followed by 360° rotation with 30-second dwell time to induce controlled annular injury.

DFP00173 (20 mg/kg) or AAV-AQP3 solution ($1×10^9$ TU/mL lentiviral vectors) were intradiscally injected using a microsyringe. Postoperative analgesia was maintained with meloxicam (1 mg/kg SC; M3937, Sigma-Aldrich, USA) for 72 h. Weekly booster injections were administered under isoflurane anesthesia. Body weights were recorded biweekly using calibrated scales (XB220A, Precisa, Switzerland).

*Micro-CT and MRI*

Following 5-week postoperative recovery, animals underwent in vivo imaging under isoflurane anesthesia (3% induction, 1.5% maintenance). High-resolution micro-computed tomography (micro-CT) scans were acquired using a SkyScan 1276 system (Bruker, Germany).

T2-weighted imaging was conducted using a 3T clinical scanner (MAGNETOM Prisma; Siemens Healthineers, Germany) equipped with a 16-channel phased-array spine coil. Sequence parameters were optimized for intervertebral disc assessment (repletion time 2000 ms; echo time 76 ms; field of view 260×320 mm; slice thickness 0.8 mm).

*Histopathologic staining*

IVD specimens were fixed in 4% paraformaldehyde (PFA; P6148, Sigma-Aldrich, USA) for 48 h at 4°C, followed by decalcification in 10% ethylenediaminetetraacetic acid (EDTA; E4884, Sigma-Aldrich, USA) solution for 28 days. Tissues were dehydrated through a graded ethanol series (70%-100%), cleared in xylene (534056, Sigma-Aldrich, USA), and embedded in paraffin (76242, Leica, Germany) using a HistoStar embedding station (Thermo Fisher Scientific, MA).

Serial 5 μm sagittal sections were cut using a RM2255 microtome (Leica) and mounted on poly-L-lysine-coated slides (S8902, Solarbio, China). Sections stained with Mayer's hematoxylin (HXG732, Baso, China) for 8 min and eosin Y (HT110232, Sigma-Aldrich, USA) for 1 min. Sections stained with 0.1% Safranin O (S8884, Sigma-Aldrich, USA) for 5 min followed by 0.02% Fast Green FCF (FCF-1, Sigma-Aldrich, USA) for 3 min. Blinded histological analysis was performed by two independent pathologists using a histological grading scale system [24].

*Statistical analysis*

All graphical representations were generated using GraphPad Prism v8.0.2 (GraphPad Software, USA). Continuous variables with normal distribution (assessed by Shapiro-Wilk test) are presented as mean ± standard deviation (SD), while non-normally distributed data are expressed as median with interquartile range. Student's-test, one-way ANOVA or two-way ANOVA analysis of variance were used to compare data from the in vitro experimental group. Non-parametric analyses were applied to ordinal histological scores and heteroscedastic datasets. Type I error rate was controlled at α=0.05 with two-tailed testing. Multiplicity-adjusted p-values are denoted as * P < 0.05 (95% CI excludes null), ** P < 0.01, ** * P < 0.001). All experiments included ≥3 independent biological replicates with technical triplicates. Randomization and blinding protocols were implemented during data collection and analysis phases.

**Results**

**Hyperosmolarity inhibited proliferation and induced apoptosis in NPCs**

The pathogenesis of IVDD has been associated with altered osmotic homeostasis. To investigate hyperosmolar effects on NPCs, primary rat NPC cultures were established and subjected to graded osmotic stress (330 and 550 mOsm/kg) for 48 h. Quantitative apoptosis assessment through immunofluorescence and flow cytometry revealed substantial cell death potentiation under hyperosmolarity conditions (550 mOsm/kg), with apoptosis rates increased compared to physiological osmolality controls (330 mOsm/kg) **(Figure 1A and B)**. This apoptotic activation was corroborated by significant elevation of cleaved caspase-3 expression, as quantified through both flow cytometric analysis **(Figure 1C)**. Mechanistic investigation via western blotting demonstrated marked upregulation of pro-apoptotic Bax protein coupled with concomitant downregulation of anti-apoptotic Bcl-2 in the 550 mOsm/kg group **(Figure 1D)**. Furthermore, ECM degradation was evidenced by significant reductions in type II collagen and the increase of MMP3 expression levels. These findings collectively establish that sustained hypertonic stress induces apoptotic cascades in NPCs while disrupting matrix homeostasis.

**Hyperosmolality promotes mitochondrial oxidative damage and ROS accumulation in NPCs**

To elucidate the mechanisms underlying hyperosmolarity-induced cellular damage, we investigated the role of reactive oxygen species (ROS) in NPCs, given their established involvement in apoptotic pathways [25]. Flow cytometric analysis revealed distinct alterations in oxidative stress markers across different osmotic conditions. Specifically, NPCs exposed to 550 mOsm/kg demonstrated a marked elevation in ROS levels **(Figure 2A)**, concomitant with a significant reduction in GPx activity **(Figure 2B)** and increased MDA content **(Figure 2C)** when compared to the 330 mOsm/kg control group.

Considering the mitochondrial origin of ROS generation, we employed dual immunofluorescence staining using Mito-SOX (red) and Mito-Tracker (green) to specifically assess mitochondrial ROS production. Intracellular and mitochondrial ROS staining showed that the Mito-Tracker (green) and Mito-SOX (red) fluorescent levels were increased in NPCs with 550 mOsm/kg **(Figure 2D)**, indicating heightened mitochondrial oxidative stress. To further evaluate mitochondrial integrity, we examined the mitochondrial membrane potential (MMP), a critical parameter of mitochondrial function, using JC-1 staining. MMP was detected by JC-1 staining which shows red fluorescence with normal mitochondria and green fluorescence in case of mitochondrial dysfunction. Hyperosmolarity increased the green fluorescence and decreased red fluorescence, indicating a decline of MMP **(Figure 2E)**. These collective findings strongly suggest that hyperosmolarity induces mitochondrial dysfunction in NPCs, leading to excessive ROS accumulation and subsequent cellular damage.

**Overexpression of AQP3 mitigated mitochondrial oxidative stress and attenuated ROS accumulation in NPCs under a high osmotic pressure condition**

AQP3, an aquaporin family member responsible for transporting water and small solutes, is notably reduced in aging IVD [23]. To investigate its role under hyperosmolarity, we assessed AQP3 expression in NPCs. RNA sequencing analysis revealed a marked downregulation of AQP3 expression in NPCs exposed to 550 mOsm/kg compared to those under 330 mOsm/kg **(Figure 3A)**. Consistent with this, Western blot demonstrated significantly reduced AQP3 protein levels under hyperosmotic conditions (**Figure 3B**), indicating osmotic stress suppresses AQP3 expression.

To evaluate the functional impact of AQP3 on NPCs survival, lentiviral AQP3 overexpression (AAV-AQP3) was performed. Flow cytometry analysis showed that AAV-AQP3-transfected NPCs exhibited a lower apoptotic rate under 550 mOsm/kg compared to controls **(Figure 3C)**. Concurrently, JC-1 staining revealed elevated MMP **(Figure 3D)**, while ROS accumulation was significantly attenuated in the AAV-AQP3 group **(Figure 3E)**. Immunofluorescence further confirmed reduced Mito-SOX fluorescence intensity in AQP3-overexpressing NPCs **(Figure 3F)**. Collectively, these findings demonstrate that AQP3 overexpression alleviates mitochondrial dysfunction, suppresses ROS production, and enhances NPCs viability under hyperosmolarity.

**Downregulation of AQP3 exacerbates NPCs apoptosis via PI3K/Akt/mTOR pathway inhibition under hyperosmolarity**

To elucidate the mechanism linking AQP3 downregulation to NPCs apoptosis, RNA sequencing and pathway enrichment analyses were conducted. To determine the role and function of NPCs from the 550 mOsm/kg group, differentially expressed genes (DEGs) analysis were performed between 330 mOsm/kg and 550 mOsm/kg and DEGs were used for gene set enrichment analysis (GSEA). GSEA revealed that pathways including PI3K/Akt and mTOR signaling pathways were significantly downregulated in NPCs with 550 mOsm/kg **(Figure 4A-C)**. Western blot confirmed reduced PI3K, Akt, and mTOR phosphorylation levels under hyperosmotic conditions. Notably, AQP3 overexpression restored their expression **(Figure 4D)**, suggesting AQP3 modulates these pathways. These results indicate that hyperosmolarity-induced AQP3 depletion promotes NPC apoptosis by impairing PI3K/Akt/mTOR pathway.

**Overexpression of AQP3 ameliorated IVDD progression in rat model**

To validate the therapeutic potential of AQP3 in vivo, an IVDD model was established via annulus fibrosus puncture (AFP). Rats received weekly injections of AQP3 inhibitor (DFP00173) or AAV-AQP3. After 5 weeks, micro-CT imaging showed narrowed intervertebral disc spaces in IVDD rats, which were partially restored by AQP3 overexpression **(Figure 5A)**. MRI analysis revealed uneven disc intensity and diminished T2-weighted signals in IVDD rats, whereas AAV-AQP3 treatment elevated T2 signals **(Figure 5B)**, indicating improved disc hydration.

Histopathological assessment demonstrated severe structural degeneration in AFP-induced IVDD rats, including collapsed disc height, NPCs loss, and fibrosis. In contrast, AAV-AQP3 administration ameliorated tissue architecture, reduced fibrosis, and lowered histological scores **(Figure 5C-E)**. Furthermore, AQP3-overexpressing rats exhibited improved weight gain compared to the IVDD group **(Figure 5F)**. These results confirm that AQP3 overexpression mitigates IVDD progression in vivo.

**Discussion**

LBP secondary to IVDD is a leading cause of global disability. This degenerative cascade is driven by multiple risk factors, including advancing age, genetic predisposition, chronic inflammation, immune dysfunction, hyperosmolarity, and nutritional deficiency. These stimuli induce apoptosis of NPCs and disruption of AF cells, leading to intervertebral space narrowing, NP herniation, and nerve root or spinal cord compression—hallmarks of chronic pain [26,27]. IVDD involves deregulated ECM synthesis, increased catabolism accompanied by cellular senescence, and apoptosis-mediated loss of nucleus pulposus cells [28-30]. Notably, microenvironmental hyperosmolarity has been identified as a critical modulator of NPCs. Studies show that exposing NPCs to 550 mOsm/kg osmolarity induces apoptosis, evidenced by nuclear fragmentation, chromatin condensation, and organelle breakdown. In contrast, physiological osmolarity of approximately 450 mOsm/kg—comparable to native NP tissue—exerts minimal apoptotic effects on cell viability [15].

Emerging evidence indicates osmotic stress-mediated AQP3 dysregulation in NPCs, with previous studies empirically demonstrating a 40% reduction in AQP3 expression at 550 mOsm/kg compared to physiological 330 mOsm/kg conditions [31]. Although osmotic sensitivity was documented, the link between AQP3 downregulation and NPC apoptosis remains unclear. We systematically evaluated apoptotic dynamics and AQP3 expression across osmotic gradients and linked these changes to ROS-mediated mitochondrial dysfunction via inhibition of the PI3K/Akt/mTOR pathway. Three principal findings emerge from this study: first, hyperosmolarity (550 mOsm/kg) induces significant AQP3 suppression, correlating with elevated intracellular ROS levels and increased NPC apoptosis. Second, pharmacological overexpression of AQP3 restores redox homeostasis—enhancing mitochondrial membrane potential stability and reducing caspase-3 activation;Third, pathway analysis reveals that hyperosmolarity disrupts the PI3K/Akt/mTOR pathway, and phosphorylated Akt levels decrease proportionally with osmotic intensity.

The pathophysiological relevance of these findings is underscored by clinical data showing 60% lower AQP3 expression in IVDD patient specimens versus healthy controls [32]. AQP3 is an aquaglyceroporin that transports both water and glycerol and also supports cellular antioxidant defenses [33]. This dual functionality explains its protective effects against $H_2O_2$-induced oxidative damage in rodent models, where AQP3 overexpression reduced lipid peroxidation and improved cell viability [34]. Our results identify AQP3 as a key integrator of osmotic adaptation, redox balance, and apoptotic signaling in IVDD pathogenesis.

Our experimental data demonstrate a distinct osmotic threshold response in NPCs, with osmotic pressure of 550 mOsm/kg inducing increased apoptosis rates and reduced proliferative capacity compared to physiological 330 mOsm/kg conditions, corroborating previous findings by Li's group regarding osmotic sensitivity in disc cells [31]. Hyperosmolarity-induced mitochondrial oxidative damage drives a cycle of ROS accumulation and cellular dysfunction. Inflammatory mediators and neovascularization-recruited inflammatory cells promote excessive ROS generation, thereby inducing oxidative stress[35]. This, in turn, damages essential intracellular macromolecules—such as lipids, DNA (including mitochondrial DNA), and proteins—ultimately accelerating cellular dysfunction and tissue degeneration[36]. Vascular endothelial growth factor (VEGF)-mediated neoangiogenesis disrupts the normally avascular NP microenvironment, potentially resulting in erythrocyte infiltration and the release of iron ions, which catalyze Fenton reactions[37]. Additionally, mitochondrial permeability transition pore (mPTP) opening facilitates cytochrome c release and activates the caspase-3-dependent apoptotic pathway[38].

Magnesium boride-alginate (MB-ALG) hydrogels have been demonstrated to promote the proliferation of senescent cells within rat IVD through activation of the PI3K/Akt/mTOR signaling cascade. Concurrently, molecular hydrogen demonstrated remarkable antioxidative efficacy by effectively neutralizing ROS, thereby protecting NPCs from oxidative stress-induced damage [39]. Sun et al. showed that hyperosmolarity suppresses the PI3K/Akt/mTOR pathway and that osteogenic protein-1 (OP-1) reverses this effect, reducing NPC apoptosis [40]. We then investigated how AQP3 downregulation contributes to NPC apoptosis under high osmotic pressure. RNA sequencing revealed that hyperosmolar conditions significantly inhibit the PI3K/Akt/mTOR signaling axis in rat nucleus pulposus cells. Importantly, adenoviral vector-mediated AQP3 overexpression successfully reversed this pathway suppression, leading to a pronounced decrease in NPCs apoptosis. These molecular interventions translated into significant improvements in the IVDD rat model, highlighting the therapeutic promise of AQP3 modulation and PI3K/Akt/mTOR pathway targeting for IVDD management. Previous investigations by Palacio-Mancheno et al. revealed elevated AQP3 expression in notochordal cell lineages following 14-day hyperosmotic conditioning of murine intervertebral disc cell cultures [20], presenting divergent outcomes from our experimental findings. This disparity likely reflects differences in osmotic stress responsiveness between notochordal cells and nucleus pulposus cells (NPCs). Comparative experiments are needed to validate this mechanism. Rodent nucleus pulposus tissue contains both NPCs and notochordal remnants—creating methodological challenges for cell isolation. Current technical constraints in distinguishing these populations through definitive molecular signatures complicate the exclusion of notochordal cell contamination during conventional NPCs isolation protocols. Developing standardized isolation methods using lineage-specific biomarkers will ensure NPC purity and reduce confounding from mixed cell populations in future studies.

**Clinical perspectives**

- Hyperosmolarity has emerged as a pathogenic driver of NPCs apoptosis during IVDD; however, the underlying mechanism remains unclear.
- Hyperosmolarity triggers NPCs apoptosis by depleting AQP3 and suppressing the PI3K/Akt/mTOR pathway, resulting in mitochondrial dysfunction and ROS accumulation.
- These findings clarify AQP3's role in IVDD and provide a molecular basis for developing AQP3-targeted clinical therapies.

**Conclusion**

This study elucidates how AQP3 downregulation mediates NPC apoptosis under hyperosmolar condition **(Figure 6)**. Our experimental show that hyperosmolarity significantly increases apoptosis of rat nucleus pulposus cells and decreases AQP3 expression. Notably, AQP3 overexpression rescues NPCs from hyperosmolarity-induced apoptosis by activating the PI3K/Akt/mTOR pathway. These findings clarify AQP3's role in IVDD and provide a molecular basis for developing AQP3-targeted clinical therapies.

**Ethics approval**

All animal handling and experimental procedures were conducted at the Guangzhou Forevergen Biosciences Laboratory Animal Center, following protocol approval by the Institutional Animal Care and Use Committee (IACUC Approval No. SYXK(Yue)2023-0186).

**Data availability**

The datasets generated and analyzed during the current study are either included in this published article or available from the corresponding author upon reasonable request. All authors have verified the availability and integrity of the data supporting this study's findings.

**Author contributions**

**Yuan Sang:** Data curation, Investigation, Methodology, Writing—review and editing. **Huiqing Zhao:** Data curation, Validation, Investigation. **Jiajun Wu:** Data curation, Formal analysis, Investigation. **Ting Zhang:** Conceptualisation, Supervision, Writing—review and editing. **Wenbin Xu:** Resources, Writing—review and editing. **Hui Yao:** Conceptualisation, Supervision. **Kaihua Liu:** Conceptualisation, Writing—original draft, Writing—review and editing. **Chang Liu:** Validation, Investigation, Methodology. **Junbin Zhang:** Validation, Investigation. **Ping Li:** Conceptualisation, Funding acquisition. **Depeng Wu:** Conceptualisation. **Yichun Xu:** Writing-review and editing. **Jianying Zhang:** Conceptualisation, Writing-original draft, Writing-review and editing. **Gang Hou:** Formal analysis, Supervision, Funding acquisition, Writing—original draft, Writing—review and editing.

**Funding**

This work was supported by the Basic and Applied Basic Research Foundation of Guangdong Province (Grant Number: 2019A1515010209 and 2023A1515011572).

**Acknowledgements**

We gratefully acknowledge the Third Affiliated Hospital of Sun Yat-sen University for providing access to their state-of-the-art research facilities.

**Competing interests**

The authors declare no conflicts of interest associated with the manuscript.

**Abbreviations**

LBP, lower back pain; IVDD, intervertebral disc degeneration; IVD, intervertebral disc; NP, nucleus pulposus; AF, annulus fibrosus; ECM, extracellular matrix; NPCs, nucleus pulposus cells; AQPs, aquaporins; AQP3, aquaporin-3; SPF, specific-pathogen-free; FBS, fetal bovine serum; RIPA, radioimmunoprecipitation assay; BCA, bicinchoninic acid; SDS-PAGE, sodium dodecyl sulfate polyacrylamide gel electrophoresis; PVDF, polyvinylidene difluoride; HRP, horseradish peroxidase; ECL, enhanced chemiluminescence; CCK-8, Cell Counting Kit-8; ROS, reactive oxygen species; DCFDA, 2',7'-dichlorodihydrofluorescein diacetate; DCF, 2',7'-dichlorofluorescein; MFI, mean fluorescence intensity; MDA, malondialdehyde; GPx, glutathione peroxidase; TBARS, thiobarbituric acid reactive substances; QC, quality control; NGS, next-generation sequencing; GO, Gene Ontology; KEGG, Kyoto Encyclopedia of Genes and Genomes; rAAV, Recombinant adeno-associated virus; TBG, thyroxine-binding globulin; ITRs, inverted terminal repeats; AFP, annulus fibrosus puncture; micro-CT, micro-computed tomography; MRI, magnetic resonance imaging; H&E, hematoxylin and eosin; SD, standard deviation; MMP, mitochondrial membrane potential; DEGs, differentially expressed genes; GSEA, Gene set enrichment analysis; VEGF, vascular endothelial growth factor; mPTP, mitochondrial permeability transition pore; NAC, N-acetylcysteine; MB-ALG, magnesium boride-alginate; OP-1, osteogenic protein-1.

**Figure legends**

**Figure 1. Hyperosmolarity inhibited proliferation and induced apoptosis in rat NPC.** (**A**) The apoptosis of NPCs was detected by TUNEL staining. Scar bar: 100 μm. (**B**) Quantitative analysis of apoptosis rates in NPCs under different osmotic pressures (330 and 550 mOsm/kg) using flow cytometry. (**C**) Flow cytometric detection of cleaved caspase-3 activation in apoptotic NPCs through intracellular staining with specific antibodies. (**D**) Western blot analysis of apoptosis-related proteins (Bax, Bcl-2) and extracellular matrix components (type II collagen, MMP3) under osmotic pressures of 330 mOsm/kg and 550 mOsm/kg. Representative immunoblots and densitometric quantification normalized to β-actin (42 kDa) show protein expression profiles. Molecular weight markers indicated in kDa. Data are expressed as mean ± SD. Significant differences between different groups are indicated as *$p<0.05$, **$p<0.01$, ***$p<0.001$.

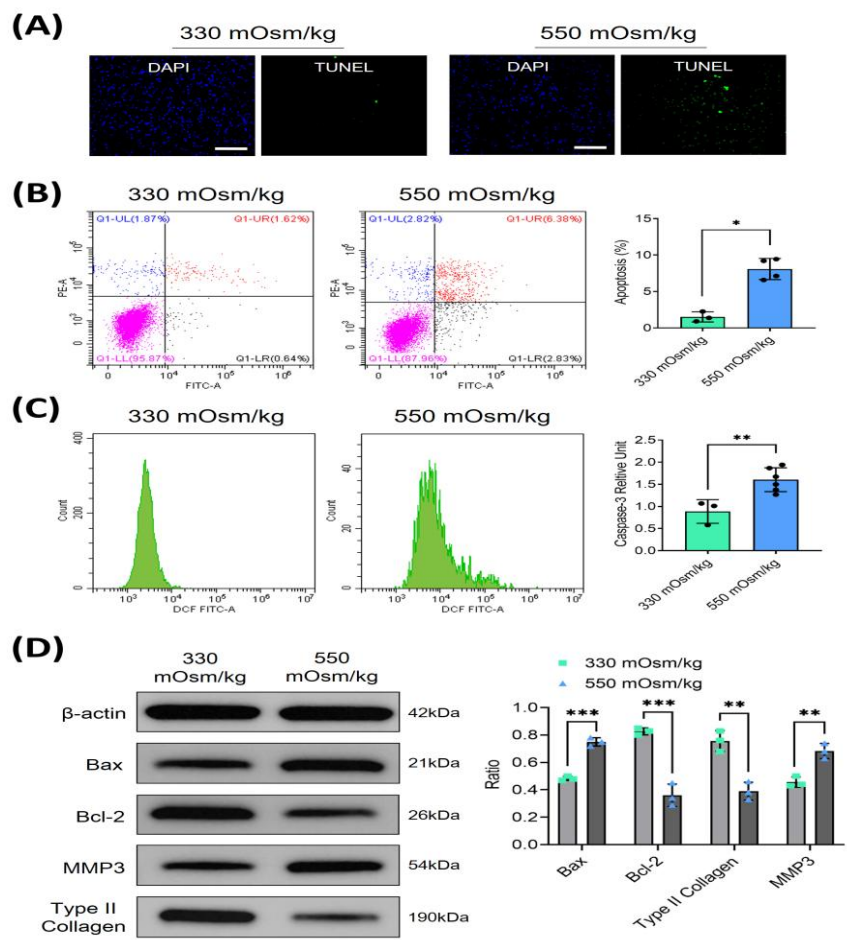

**Figure 2. Hyperosmolality induces mitochondrial oxidative damage and ROS accumulation in NPCs.** (**A**) Flow cytometric quantification of intracellular ROS levels under osmotic pressures of 330 mOsm/kg and 550 mOsm/kg. (**B**) The activity of GPx measured by colorimetric assay showing enzymatic antioxidant capacity attenuation under different osmotic pressures (330 mOsm/kg and 550 mOsm/kg). (**C**) MDA content quantification via thiobarbituric acid reaction, reflecting lipid peroxidation levels under different osmotic pressures (330 mOsm/kg and 550 mOsm/kg). (**D**) Confocal imaging of mitochondrial superoxide (MitoSOX, Red) and mitochondrial morphology (MitoTracker, Green) co-localization. Scar bar: 200 μm. (**E**) MMP assessment using JC-1 fluorescence shift (red/green ratio). Scar bar: 100 μm. Data represent mean ± SD from three biological replicates. Significant differences between different groups are indicated as *$p$<0.05, **$p$<0.01.

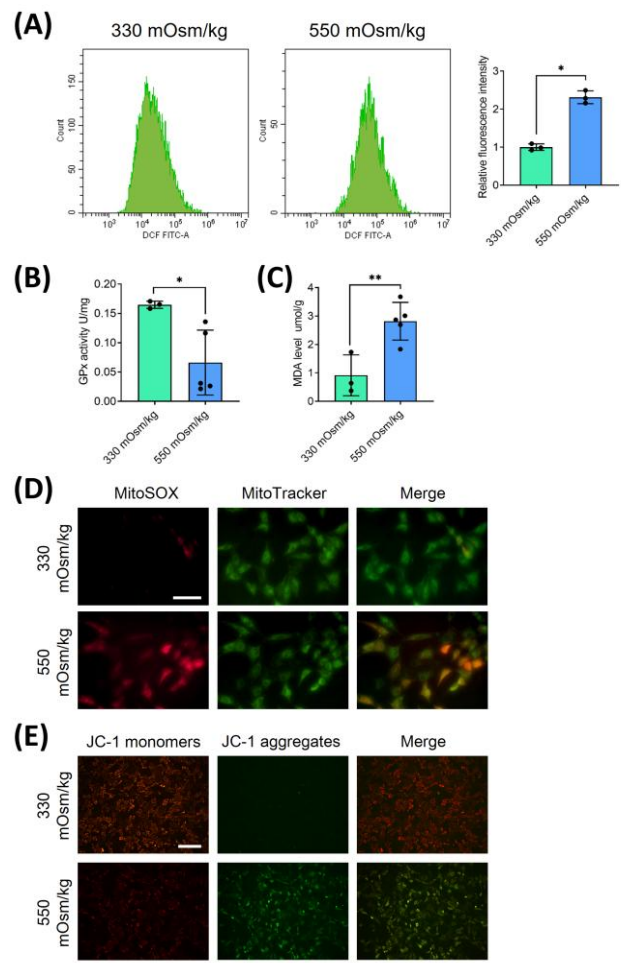

**Figure 3. Overexpression of AQP3 mitigated mitochondrial oxidative stress and attenuated ROS accumulation in NPCs under a high osmotic pressure condition.** (**A**) mRNA expression of AQP3 in NPCs under different osmotic pressures (330 mOsm/kg and 550 mOsm/kg). (**B**) Western blot analysis of AQP3 under different osmotic pressures (330 mOsm/kg and 550 mOsm/kg). Representative blots and densitometric quantification normalized to β-actin (42 kDa) are shown. (**C**) The quantitative statistics of apoptotic NPCs. (**D**) MMP preservation assessed by flow cytometry in NPCs of each group. (**E**) Intracellular ROS levels measured using flow cytometry. (**F**) Confocal microscopy visualization of mitochondrial superoxide (MitoSOX, Red) and mitochondrial morphology (MitoTracker, Green) co-localization. Scar bar: 200 μm. Data represent mean ± SD from three biological replicates. Significant differences between different groups are indicated as \*$p<0.05$, \*\*\*$p<0.001$.

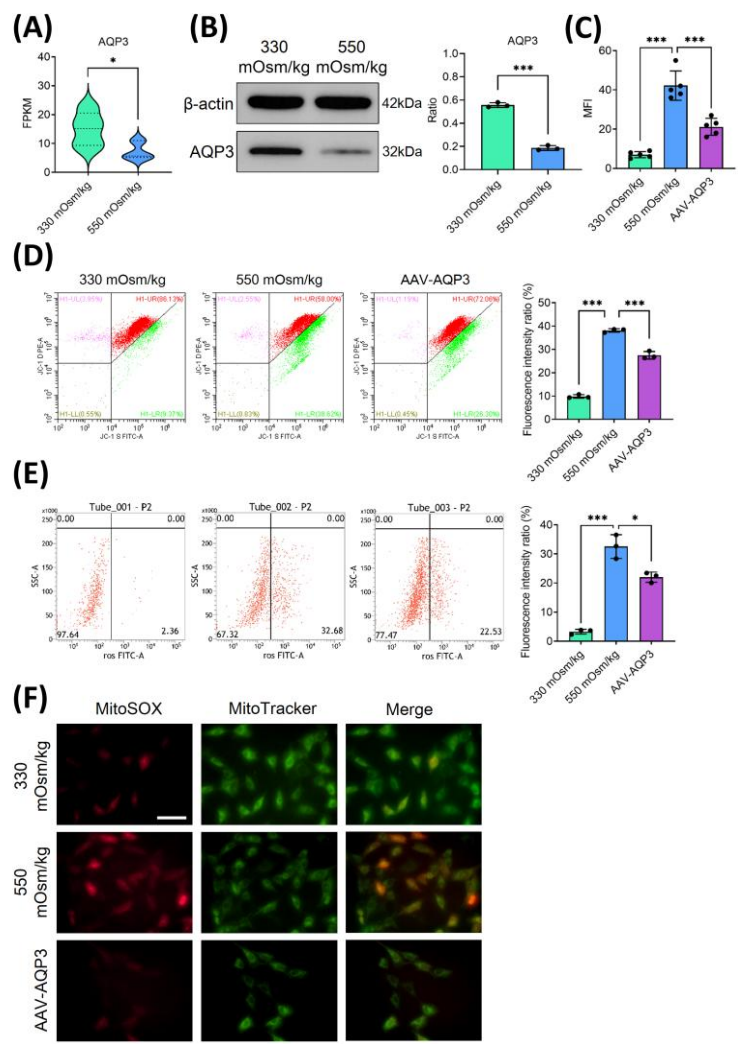

**Figure 4. Downregulation of AQP3 exacerbates NPCs apoptosis via PI3K/Akt/mTOR pathway inhibition under hyperosmolarity.** (**A**) GSEA of selected enriched KEGG pathways for NPCs in 550 mOsm/kg group compared with 330 mOsm/kg group. Spot diameter indicates gene count. Colour represents the log10-transformed p value. (**B**) Enriched KEGG pathways of the genes downregulated in 550 mOsm/kg group compared with 330 mOsm/kg group for NPCs. (**C**) GSEA demonstrating coordinated suppression of PI3K/Akt and mTOR signaling components. The y-axis represents enrichment score, and the x-axis denotes genes represented in gene sets. (**D**) Western blot analysis of p-mTOR, p-PI3K and p-Akt level in rat NPCs of each group. Data represent mean±SD from three biological replicates. Significant differences between different groups are indicated as **$p<0.01$, ***$p<0.001$.

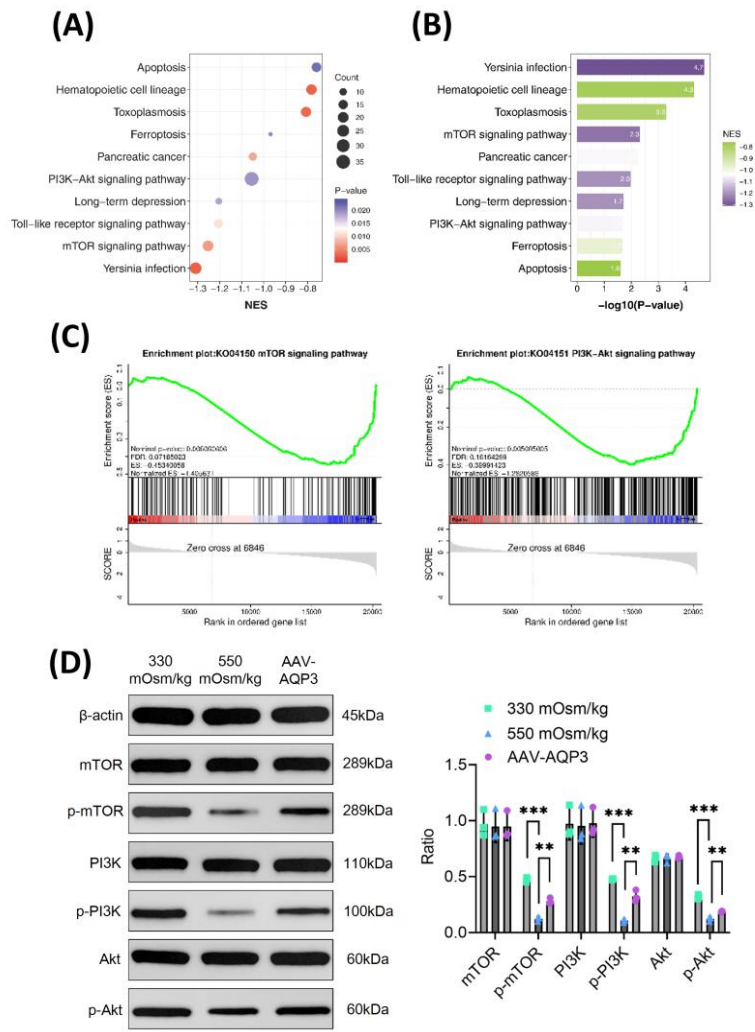

**Figure 5. Overexpression of AQP3 ameliorated IVDD progression in rat model.** (**A**) Schematic representation of experimental design: Control (n=6), AFP (n=6), AFP+DFP00173 (n=10), and AFP+AAV-AQP3 (n=6) groups. Interventions were administered via intradiscal injection weekly for 5 weeks post-puncture. Micro-CT sagittal reconstructions demonstrating intervertebral space height preservation. (**B**) T2-weighted MRI axial sections. Operated levels (Co6/7) are demarcated by white triangles. (**C-E**) Histopathological evaluation of NP tissues integrity. Representative H&E staining and Safranin-O/fast green staining with semi-quantitative histomorphometric analysis. Scale bar: 500 μm. (**F**) Average growth weight curves of rats in each group. Data represent mean±SD. Significant differences between different groups are indicated as *$p$<0.05, ***$p$<0.001.

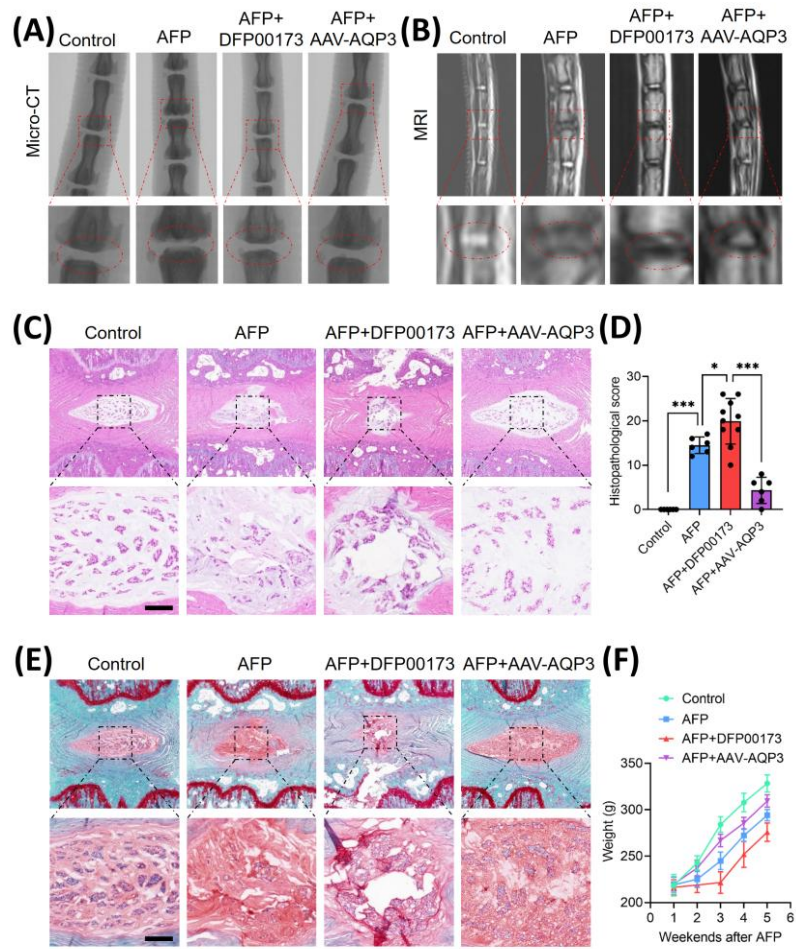

**Figure 6.** Schematic illustrating the mechanism of downregulation of AQP3 inducing NPCs apoptosis through PI3K/Akt/mTOR pathway suppression in a high osmotic pressure environment.

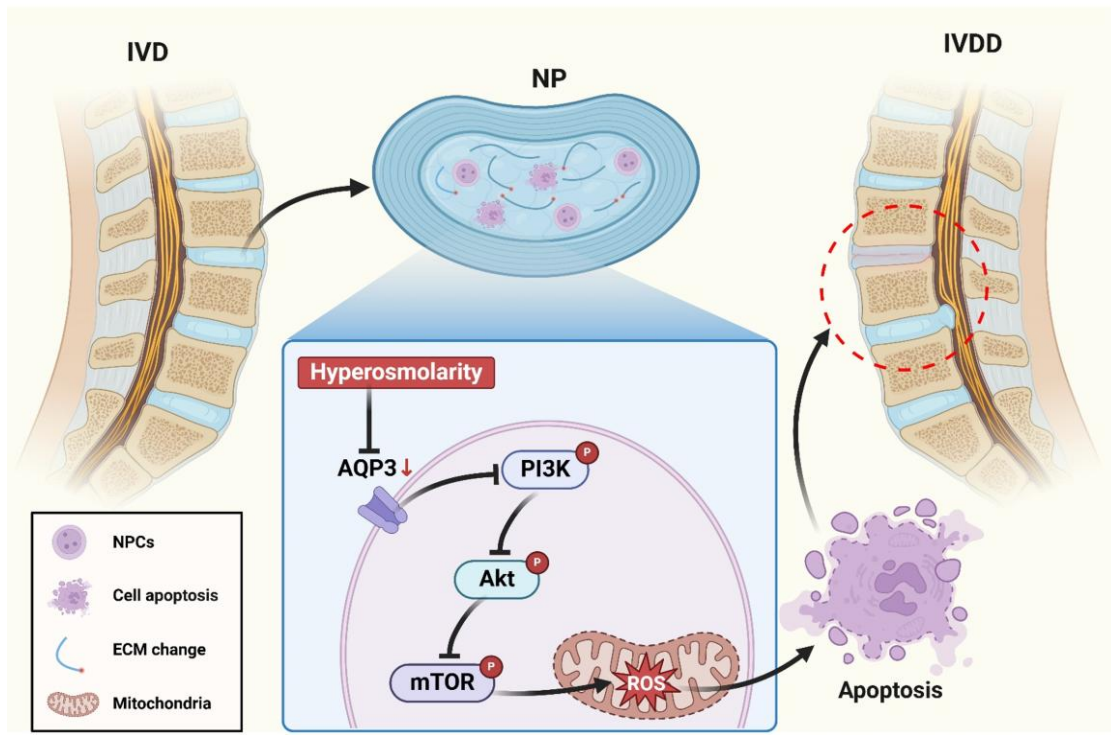